# iCare: A Mobile Health Monitoring System for the Elderly

Ziyu Lv, Feng Xia, Guowei Wu, Lin Yao, Zhikui Chen
School of Software
Dalian University of Technology
Dalian 116620, China
e-mail: f.xia@ieee.org

*Abstract*—This paper describes a mobile health monitoring system called iCare for the elderly. We use wireless body sensors and smart phones to monitor the wellbeing of the elderly. It can offer remote monitoring for the elderly anytime anywhere and provide tailored services for each person based on their personal health condition. When detecting an emergency, the smart phone will automatically alert pre-assigned people who could be the old people's family and friends, and call the ambulance of the emergency centre. It also acts as the personal health information system and the medical guidance which offers one communication platform and the medical knowledge database so that the family and friends of the served people can cooperate with doctors to take care of him/her. The system also features some unique functions that cater to the living demands of the elderly, including regular reminder, quick alarm, medical guidance, etc. iCare is not only a real-time health monitoring system for the elderly, but also a living assistant which can make their lives more convenient and comfortable.

*Keywords-healthcare; assisted living; health monitoring; body sensor; medical guidance*

## I. INTRODUCTION

The aging people in the contemporary society have brought great pressure and many seniors have been living alone without anyone accompanied because their offspring are busy with work and have to struggle with severe competition. For those seniors who live independently in their own homes, there is an increasing risk of falls and strokes which could threaten their lives. A lot of money and research effort has been spent on making people aware of the warning signs [1]. Getting the elderly to recognize the warning signs is not easy. And it indicates that people who have had a heart attack have a sudden death rate that is 4 to 6 times that of the general population [2]. The New England Journal of Medicine draws a conclusion that the chances of surviving a fall, heart attack or stroke are six times greater if the elderly get help within an hour. Therefore, the elderly with no one accompanied need real-time monitoring to reduce the anxiety of them and the risk of accidents. However, the financial and staffing cost of caring for the increasing numbers of aged persons in nursing homes or hospitals will be a huge challenge. In this background, a very worthy and challenging issue is how to provide the elderly with the real-time, long-term and nonintrusive assisted living and remote healthcare services.

With the development of the technologies such as mobile computing, distributed computing and wireless sensor network, it is possible to provide the elderly with healthcare services that can monitor the elderly anytime anywhere. Various wireless communication motes cooperate with medical sensors to support healthcare and monitor people's health, especially the elderly suffering from diabetes, high blood pressure or heart disease [3]. For instance, in Australia, Pro Medicus makes a success of IT Healthcare [4]. It develops one secure intelligent system for transmitting medical results to doctors. Wireless sensor network provides the useful method to remotely acquire and monitor physiological signals without the need of disrupting the patient's normal life [5-9].

We intend to develop a mobile health monitoring system, called iCare. It can monitor the old people anytime and anywhere. Through vital physiological data monitoring, accidents perception, real-time emergency response and other functions, our system will reduce sudden accidents and life-threatening [10, 11]. It is convenient that the elderly can have access to medical care at home. At the same time, we feature auxiliary functions which cater for the need of the elderly without anyone accompanied or health professionals as their life assistant. Therefore, the system iCare acts as not only one remote health monitoring system, but also the life assistant. Moreover, it will help establish the health insurance system and contribute to the aging society.

The rest of the paper is set out as follows. Section 2 outlines related work. The requirements and system design of the system iCare are described in Section 3 and Section 4 respectively. Section 5 illustrates the prototype of iCare and shows user interface. Section 6 concludes this paper.

## II. RELATED WORK

Today, personal healthcare is one of emerged areas of research. Rodriguez *et al* [12] have made a classification which divides the solutions into three groups. The first group records signals and takes action off-line. The second group has the feature that systems perform remote real-time processing. The last group provides local real-time processing, with taking into account the level of mobility.

The Holter device that records patients' ECG for 24 to 48 hours is analyzed afterwards by doctors. Therefore, Holter belongs to the group one. The drawback of Holter devices are their deficiency in offering real-time monitoring and no immediate action when the accident occurs. In order to overcome the restriction, many systems and devices are developed. Vitaphone commercializes a card that can transmit ECG data to a mobile phone. The mobile phone automatically transmits ECG data to the service center where ECG data are analyzed [13]. Similarly, Cardio Control [14], MediSense [15] and MobHealth project are all included in the group two, using mobile phone/PDA to get physiological signals and sending signals to other devices in which physiological signals are remote real-time monitored. Besides, MORF is also one respective application of the second group. It uses mobile phone as an intermediary to get vital data from various sensors and transmit data to the server which processes the data [16]. However, the above applications still present certain limitations related to the fact that the analysis is not performed in the place where the signal is acquired. There may be a loss of efficiency in the wireless network when physiological signals are sent.

Compared with the group two, the third group performs the local real-time monitoring in order to detect some anomalies and send alert to a control center or a hospital. Wu et al [17] proposes a wearable personal healthcare and emergency aid system called WAITER. It employs tiny wearable sensors to continuously collect users' vital signals and uses Bluetooth devices to transmit the sensory data to a mobile phone, which can perform on-site vital data storage and processing. After local data processing, the mobile phone can periodically report users' health status to the healthcare centre via its GSM module and issue alert for medical aids when detecting the emergency. But it only develops one relatively static monitoring system in which the status is set statically and doctors are called when mobile phone send alert messages. It is not sufficient in real-time and dynamic monitoring. Prognosis [18] is a physiological data fusion model of wearable health-monitoring system for people at risk which contains decision support system and finite-state machine. It can estimate users' health status and offer corresponding alerts. Gay and Leijdekkers [19] have developed one application that can monitor the wellbeing of high risk cardiac patients using wireless sensors and smart phones. Depending on the situation, the smart phone can automatically alert pre-assigned caregivers or call the ambulance. Although it performs real-time monitoring, it still does not consider the increasing life demand of the elderly. And it only provides real-time monitoring rather than real-time healthcare services.

Our system is also a mobile health system for the elderly with tailored services depending on their personal health condition. However, it is not only a remote and real-time monitoring system which takes both doctors and the old people's family and friends into account, but also a life assistant of the elderly including unique auxiliary functions. The unique auxiliary functions cater to the increasing life demand of the elderly who are with nobody accompanied. It also integrates both the health information system and medical guidance to assist the life of the elderly better. Therefore, it provides convenient and indispensable healthcare service for the elderly.

III. REQUIREMENTS SPECIFICATION

Our system is mainly based on the current real-time health monitoring of old people, catering for the demand of assisted living and the development of healthcare. We will design one mobile health monitoring system called iCare in order to monitor personal health of the elderly anytime and anywhere. At the same time, the elderly not only require real-time health monitoring, but also call for a life assistant due to their physical status of discomfort and decline in life skills (e.g. visual and mobility decline, memory loss). Therefore, iCare also offers unique auxiliary functions which are advanced when the daily demand taken into account for the elderly as the life assistant. Moreover, we also provide the personal health information system and the medical guidance which contains the communication platform and medical knowledge database. Therefore, the main functions of our system are described below.

- A real-time mobile health monitoring system for the elderly to monitor the health condition anytime and anywhere.
- The personal health information system.
- The medical guidance.
- A living assistant that offers some auxiliary functions to cater for the living demands of the elderly.

*A. Mobile Health Monitoring System*

iCare is mainly a real-time mobile health monitoring system for the elderly. With the development of mobile computing and wireless sensor technology, combining smart phone, body sensors and web technology can offer real-time mobile health monitoring for the elderly. Our system will use body sensors and devices to collect physiological signals from the elderly and transmit them to smart phone that will process physiological data locally and alarm automatically to the emergency centre and pre-assigned people when data exceed the threshold of the fixed device. The emergency centre will call an ambulance to the current location of the elderly. The location information can be gained from the alarm message sent to the emergency centre. Sensors and devices can be tailored depending on the old people's health condition. The physiological data will be not only analyzed locally, but also sent to the server in bulk to construct the personal health information system. With viewing the current and history condition of the old people in the personal health information system, doctors remotely set the thresholds and give advices which can guide the old people to adjust themselves to health mode.

*B. Personal Health Information System*

Our system can act as the personal health information system on the server. The smart phone gets physiological data from sensors, and locally stores physiological data and other information such as alarm message, advice message. Smart phone will send those physiological data and other information in bulk periodically to the server and the server stores them in database to construct the personal health information system. Doctors, family and friends of the old people who should be granted first can access to viewing the current and history condition of their subjects. Therefore, doctors can know the condition of the elderly and alter related thresholds so that dynamic monitoring can be guaranteed.

### C. Medical Guidance

The system can offer the medical guidance. It can provide both one communication platform and the medical knowledge database for doctors and other users including the elderly, their family and friends. Through the medical guidance, they can obtain medical knowledge and know related measures immediately. That family and friends of the elderly cooperate with doctors to monitor and assist the elderly through the medical guidance can become the essential part of our system. It not only makes users gain more knowledge in the aspect of healthcare, but also provides self-care guide service.

### D. Living Assistant

In order to cater to the life demand of the old people, the system iCare will offer unique auxiliary functions as the life assistant, which include regular reminder, quick alarm, location tracking, etc. Considering that the memory power of the old people is poor, regular reminder can remind the elderly to take medicine at certain time and give some advices according to the current climate situation. In order to make the elderly alarm quickly when they feel uncomfortable themselves, quick alarm can offer the quick-button for immediate alarm. Location tracking is used to get real-time location of the elderly so that they can be found immediately when detecting emergency. The functions above can play a vital role in the life of the elderly.

## IV. SYSTEM DESIGN

This section describes the design of the system iCare based on the requirements. Fig. 1 shows the architecture of our system iCare.

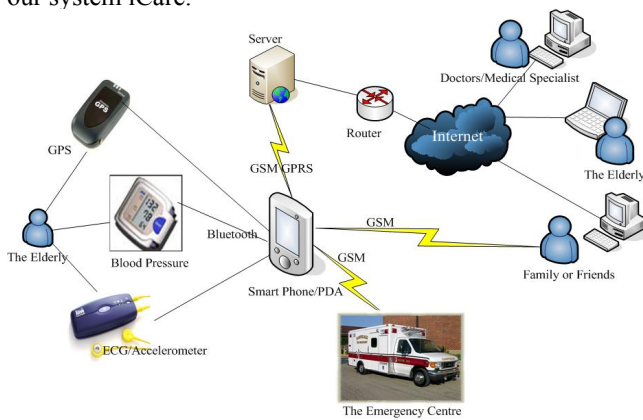

Figure 1. System architecture.

The system can be divided into four parts. The first one is devices, including different body-sensors and other medical devices. Then, the second part is smart phone that plays a vital role in our system as an intermediary. It receives physiological data from sensors, processes them and transmits them to the server. Smart phone will monitor physiological data get from sensors and automatically alert to the emergency centre, family and friends of the elderly when detecting the emergency. In addition, it also designs unique auxiliary functions as a life assistant. The server is the third one that acts as the personal health information system. In the meanwhile, it also plays the role of the medical guidance which can offer the real-time medical guidance for users. The last part contains the emergency centre that will receive alarm messages via GSM protocol and call an ambulance to the location of the subject in the emergency status.

### A. Sensors and Devices

Sensors and medical devices are necessary to collect physiological data from the elderly in our system iCare. Recently, cardiac disease is very common in the elderly who are at high risk. For this reason, ECG signal is the obvious data that needs to be collected continuously and should be given priority over all other sensor data. At the same time, high blood pressure is another important risk factor for diseases [2] and regular monitoring is essential. According to National Heart Foundation of Australia, the physical exercise improves the live expectancy. Thus, evaluating the level of activity of the elderly can be an important indication, which can use an accelerometer to achieve. Moreover, being overweight or obese should also be concerned because it may lead to high risk disease. Therefore, in our system we could use e.g. an integrated Bluetooth ECG/Accelerometer sensor from Alive Technologies [20] and a Bluetooth enabled blood pressure [21]. In addition, our system provides tailored services which allow users to choose sensors and devices depending on the personal health condition. The configuration of sensors can also be set by their doctors.

### B. Smart Phone

Smart phone should communicate with sensors and the server, and send alarm messages to the emergency centre and family of the subject when the elderly detected in the state of emergency. It also contains some auxiliary functions such as regular reminder, quick alarm and medical knowledge inquiry as a life assistant of the elderly. Therefore, the application of smart phone is complicated and plays a vital role in our system. It is divided into nine modules. The nine modules include data receiving, data processing, data sending, basic information setting, configuration, inquiry, quick alarm, location tracking and reminder. Data receiving is used to receive data from sensors and the server, and Data sending is the interface to send messages to the server, the emergency centre and preset persons. Basic information setting and configuration are used to set basic information of all users and configuration of functions on smart phone. Inquiry module allows the elderly to check alert records and advices sent by doctors. Quick alarm provides the elderly with quick button to alert to the emergency centre. The location of the subject can be got and recorded through location tracking module that is also essential in the alarm function. Reminder will be used to reminder the elderly of trivial affairs and medical events. The main module will be described below.

*1) Data receiving and data sending:* Data receiving module is designed for receiving data from sensors and the server. Bluetooth is used when smart phone communicates with sensors. Sensors collect physiological data from the old people. Then, smart phone can get physiological data from sensors via Bluetooth. Smart phone receives advice and threshold information from doctors in the form of SMS via GSM. Especially, data receiving module contains data-formatted function that formats data and classifies them into three different categories (threshold, advice and physiological data), which must be done before data processing module. In addition, data sending module also uses Bluetooth for communicating with sensors. It uses socket interface to transmit data to the server in bulk via GPRS. Moreover, smart phone will send SMS messages to the emergency centre, family and friends of the elderly when generating alarm information.

*2) Data processing:* Data processing module is to process data depending on different categories of data which are get from data receiving module and take related action. For physiological data, the monitor function will be called. The monitor function which belongs to the data processing module is the main function that can provide real-time monitoring for the elderly in our system. Firstly, it stores the receiving data in the record store, updates the old data and notifies the elderly. Then, the value of data should be monitored and compared with the fixed threshold that is preset by doctor remotely. If it exceeds the threshold, the alarm flag will be marked. When the flag is marked twice in the adjacent monitoring, our system will call the alarm mechanism. The alarm function will ask the subject whether to cancel the alarm because ambient interference may generate false alarm. If the subject cancels the alarm within the alarm waiting time that is preset in configuration module, iCare will return to normal mode and continue to monitor data. Otherwise, when there is no action taken within the alarm waiting time or the elderly confirm the alarm, our system will automatically send alarm messages to the emergency centre and the targets preset in family and friends of the subject. In the monitor function, location tracking is called for getting the location of the subject which will be included in the alarm information. Additionally, the system receives threshold and advice from doctors remotely so that real-time dynamic monitoing is guaranteed. When receiving threshold information, it will store the information, update the threshold of the related record and notify the elderly the arrival of the new threshold. Similarly, if the received data is advice, the system stores the advice, notifies the elderly of the arrival of new advice and inquires about whether the elderly wants to cheek the new advice.

*3) Configuration and basic information setting:* The two modules are basic modules in our system. The basic information setting module is just to record basic information of users that include the elderly, doctors, family and friends of the elderly. The module configuration is used for initial and basic configuration of the system, including system mode, device selecting, transmitted time setting and alarm target setting. Transmitted time setting have two units: alarm waiting time that is checked when it is time to inquiry whether current situation is real alarm in the alarm function of the monitor module. And transmitted time is the interval when physiological data should be transmitted in bulk to the server.

*4) Auxiliary function:* Considering the demand of the life assistant of the elderly, some auxiliary functions are designed on smart phone. They are inquiry, location tracking, quick alarm and regular reminder. Inquiry can enable users to check their alarm records and advice of doctors. Especially, medical knowledge inquiry is designed, which can enable users to learn any medical knowledge they want to know or knowledge related to their diseases. The special function is unique compared with other similar health monitoring system. Location tracking is designed on the phone side. It is called to get current location of the subject which will be added in alarm messages when alarming to the emergency centre and the subject's family and friends preset. Combining the technology of location tracking and GPS develops this module, which can provide a rough indication of the location of the subject both indoors and outdoors. Quick alarm and regular reminder are also the unique features that cater to the demand of the elderly. Quick alarm provides quick button that can cause alert to the emergency centre immediately when the elderly feel comfortless, which is very convenient for the elderly without anyone accompanied. Considering that the memory power of the elderly is dropping, regular reminder is designed. Currently, we design two reminders (medicine reminder and climate reminder) in our system. Medicine reminder is designed to reminder the elderly to take medicine on schedule. Users can set the frequency to reminder themselves depending on different medicine. We can design three modes for medicine reminder: every six hours, every eight hours and every twelve hours. Another reminder (climate reminder) is designed to tell the condition of climate and advise them to take action. It uses climate proxy server to get real-time weather information and design climate-advice knowledge database to offer corresponding advices. There are also three modes for climate reminder, including every one day, every two days and every three days. When the reminder is enabled, it will give current weather conditions and a kindly reminder.

*C. Server*

The server not only acts as the personal health information system in which physiological data and other information can be stored and users can check all kinds of records, but also serves as the medical guidance. The server in our system iCare can be seen as the remote side where doctors can use the web application and database on the server to monitor the elderly and give related advice remotely. Therefore, the server application can be divided into two modules including the personal health information system and the medical guidance.

*1) Personal health information system:* The personal health system can store physiological data and other information in the database on the server and show current and history condition with the way of comparative. As the server should communicate with smart phone, it has similar module: sending and receiving module. Sending module designs the function that can send thresholds and advice of doctors to smart phone in the form of SMS via GSM. And the server use socket interface to receive physiological data and other information from smart phone in bulk. The roles of users contain doctors, the elderly as well as family and friends of the elderly who should be granted before. The elderly can check their own condition via the Internet. They can grant their family and friends in order to give different limits of authority. The group of family and friends should be granted by their subjects. They can have the same function as the elderly when they are granted. Doctors are the most important users on the server. Except that they can check physiological data in order to know the condition of the elderly, doctors can also update related threshold and give real-time advice by sending SMS messages to smart phone.

*2) Medical guidance:* The function is mainly catering to the increasing demand of assisted living. It includes the communication platform and the medical knowledge database. The communication platform can allow doctors and other users to communicate with each other. The elderly, their family and friends can leave messages which can contain medical questions and related advice for doctors. Then, doctors can check the messages and give related replies. Moreover, doctors can give some advices for family and friends of the elderly when they view the current condition of the elderly, which makes it possible that family and friends of the elderly can cooperate with doctors to take care of the elderly. We also design to establish the medical knowledge database. It can guide common medical knowledge for users, and show related advices and measures for people who are in the emergency state but without anyone accompanied or health professionals nearby. Medical specialists play a vital role in building the medical knowledge system. We offer medical specialists the right to add medical knowledge of their area. And the other group consisting of excellent medical specialists has access to evaluating the existing medical knowledge, which can form the confidence level of the evaluative criteria. Currently, we design three confidence levels: credit, general, weak. When the elderly, their family and friends face the emergency or medical problem, they can inquire about medical knowledge and related measures to solve the case immediately. According to the keyword and area, the knowledge database will return related medical records with their confidence level. The medical records can be sorted by the confidence level. In fact, the medical records of weak level will not be recommmanded. Users have the right to set minimum level to filter results. Moreover, the evaluative criteria should be dynamically altered depending on medical specialists' verification and the response of users.

*D. Emergency Centre*

The emergency centre can offer the emergency treatment immediately when it receives the alarm message. The alarm message that contains the location information is in the form of SMS. Therefore, the emergency should develop one application which can receive SMS message and process messages to get the current location of the subject. Then it calls an ambulance to the current location of the subject.

V. PROTOTYPE IMPLEMENTATION

Our system iCare acts as one mobile health monitoring system for the elderly, with taking into account unique auxiliary functions as the life assistant. Although it has been divided into four parts in design section and every part has different framework and platform, they are integrated by the bend of heterogeneous network. The application executed on smart phone is developed in Java ME with MVC structure. It implements the real-time health monitoring and some unique auxiliary functions such as regular reminder, quick alarm and location tracking. The web application on the server is developed in JSP. It not only acts as the personal health information system, but also serves as the medical guidance. In the first stage of prototype, we mainly focus on the main user interface and unique cases of our project.

*A. Main Interface*

The application on smart phone designs the real-time health monitor as well as other unique functions as the life assistant. Fig. 2 shows the main interface of smart phone.

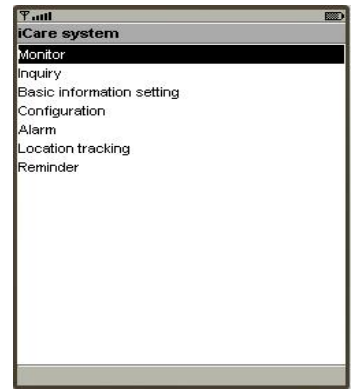

Figure 2. Main interface on smart phone.

Users can choose the items to enter related functions. Before it works, it is essential to input the basic information of the user, his doctors, the user's family and friends. And configuration is also necessary, e.g. choosing devices for tailored service, setting phone numbers for alerting, etc.

## B. Real-time Health Monitoring

The real-time health monitoring can monitor real-time condition of the elderly and automatically alert to the emergency centre and pre-assigned people. When real-time health monitor starts, it receives physiological data from sensors and processes them. Fig. 3 shows the case of the real-time health monitoring. It displays the real-time physiological data and the current threshold value. Only ECG has data because only ECG is chosen in configuration. When detecting the emergency, it automatically alert to the emergency centre and pre-assigned people. The alarm message received by the emergency centre contains current time, the identification number of the elderly, the identification number of sensor as well as the location information of the elderly. The function should receive the thresholds of physiological signals and advice from doctors remotely via SMS so that the system offers dynamic monitoring. From Fig. 4, we can see the advice sent by doctors to give real-time medical guidance.

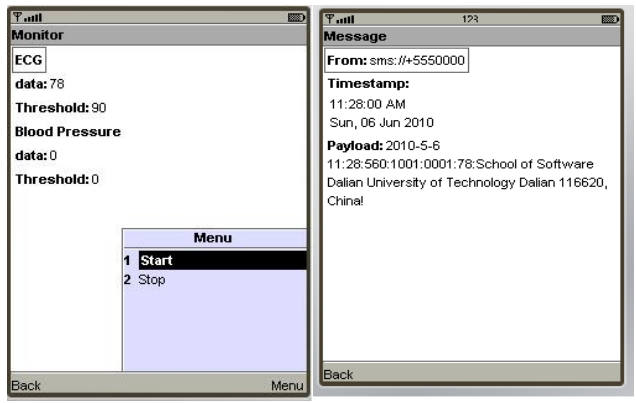

Figure 3. Real-time health monitoring.

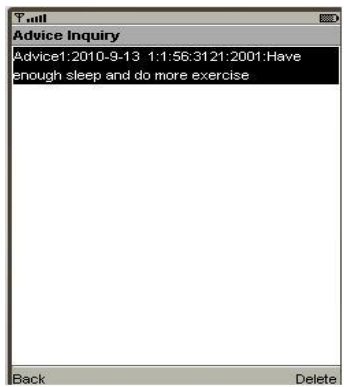

Figure 4. Advice interface.

## C. Regular Reminder

Considering that the memory power of the old people is dropping, one unique function, regular reminder, is designed as the life assistant. Currently, we implement two reminders: medicine reminder and climate reminder. When the reminder is enabled and the period is set, it will periodically remind the old people. Fig. 5 shows the case of climate reminder. The climate reminder can display the current climate condition and give a kindly reminder.

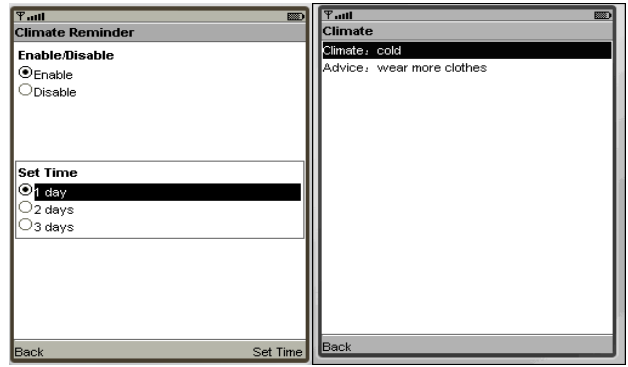

Figure 5. Weather reminder

## D. Medical Guidance

The current version of iCare provides a communication platform and the medical knowledge database on the server, which facilitate the medical guidance. The medical knowledge database allows users to learn medical knowledge and know what action should be taken in case of emergency. When users fill in keyword and area, the medical knowledge database can return related measures and advice with certain confidence level. Currently, the basic functions of the medical database have been implemented. The evaluative criteria are still in the test and improvement phase with the efforts of medical specialists. Using the communication platform, the elderly, their family and friends can send messages to doctors and seek advice. Doctors can be informed of the status of the elderly and current problem immediately, and hence give advice remotely.

## VI. CONCLUSION

We designed a mobile health monitoring system for the elderly, called iCare, which can not only dynamically monitor the elderly anytime anywhere and automatically alarm to the emergency centre in the emergency situation, but also play a role in acting as a living assistant. It provides auxiliary functions as the living assistant, including e.g. regular reminder, quick alarm. At the same time, iCare also acts as the personal health information system which allows doctors to view current and history condition of the elderly, set thresholds for sensors and give advices remotely, which is the essential part for tele-monitoring of the elderly. Additionally, the medical guidance that includes the communication platform and the medical knowledge database is designed to serves as the real-time medical guidance for the elderly, which is unique compared with other monitoring systems. Moreover, our system takes into account the role of family and friends, which make it possible that they can cooperate with doctors to take care of the elderly better. Therefore, the system iCare will not only play an essential role in assisting living, but also bring the development in healthcare.

The system is in the first stage. We have implemented basic functions of our system. A lot of work should be done in order to further improve the current system. The future work will mainly focus on the following aspects. Firstly, the user interfaces of the application are rough. We will make user interfaces more friendly and predictable depending on the characteristics of the elderly. In addition, the function as the medical guidance should be improved, especially the medical knowledge database. Similarly, it may have knowledge database bottle-neck even if the medical knowledge records are acquired from medical specialists and the evaluation mechanism is designed. We tend to use other artificial intelligence technology to optimize and improve this function. Thirdly, we attempt to propose the idea of the regional emergency service while only the emergency centre is designed in the first stage. When detecting the emergency of one subject, the system will alert to the regional emergency centre to which the subject belongs. As the regional emergency is near the place of the subject and it realizes the task-sharing of the emergency centre, the rescue work can be more efficient. However, there are two prerequisites for the regional emergency service. The location tracking function should be improved and the regional emergency service relies on perfect community service. Furthermore, more unique functions catering to the demand of the old people will be explored because the uniqueness of iCare mainly focuses on the assisted living functions.

ACKNOWLEDGMENT

The authors thank Qiqi Fan, Jie Hu, and Donglu Wang at Dalian University of Technology for their help in programming. This work was partially supported by the National Natural Science Foundation of China under Grants No.60703101 and No. 60903153, and the Fundamental Research Funds for the Central Universities.